\documentclass[12pt]{article}
\usepackage{amsfonts,amssymb,epsfig,amsmath}
\usepackage{diagbox,cite}

\renewcommand{\baselinestretch}{1.2}

\def\det{{\rm det}}

\newcommand{\be}{\begin{eqnarray}}
\newcommand{\ee}{\end{eqnarray}}

\newcommand{\bn}{\begin{enumerate}}
\newcommand{\en}{\end{enumerate}}

\begin{document}

\makeatletter \@addtoreset{equation}{section} \makeatother
\renewcommand{\theequation}{\thesection.\arabic{equation}}
\renewcommand{\thefootnote}{\alph{footnote}}

\begin{titlepage}

\begin{center}

\vspace{2cm}

{\Large\bf F-Theory and {\cal N}=1 SCFTs in Four Dimensions}

\vspace{2cm}

\renewcommand{\thefootnote}{\alph{footnote}}

{\large 
David R. Morrison* and Cumrun Vafa$^{\dagger}$}

\vspace{0.7cm}

\textit{*Departments of Mathematics and Physics, University of California Santa Barbara, CA 93106, USA}

\textit{ $^\dagger$Jefferson Physical Laboratory, Harvard University, Cambridge,
MA 02138, USA}\\

\vspace{0.7cm}

\end{center}

\vspace{1cm}

\begin{abstract}

Using the F-theory realization, we identify a subclass of 6d (1,0) SCFTs whose compactification on a Riemann surface leads to ${\cal N}=1$ 4d
SCFTs where the moduli space of the Riemann surface is part of the moduli space of the theory.   In particular we argue that for a special case of these theories (dual to M5 branes probing ADE singularities),
we obtain 4d  ${\cal N}=1$ theories whose space of marginal deformations is given by the moduli space of flat ADE connections
on a Riemann surface.  
\end{abstract}

\end{titlepage}

\renewcommand{\thefootnote}{\arabic{footnote}}

\setcounter{footnote}{0}

\renewcommand{\baselinestretch}{1}

\tableofcontents

\renewcommand{\baselinestretch}{1.2}
\section{Introduction \label{sec:INTRO}}
Motivated from F-theory constructions, recently a classification of 6d SCFTs has been proposed \cite{Heckman:2015bfa} (see also \cite{Bhardwaj:2015xxa}). It is natural to ask how from this classification we get new theories in lower dimensions.  A case of particular interest is 4 dimensions.  SCFTs in 6d can have ${\cal N}=(2,0)$ or ${\cal N}=(1,0)$ supersymmetry.  If we start with the $(2,0)$ theory we obtain 4d theories with ${\cal N}=4,2,1$ depending on whether we use $T^2$ or a Riemann surface whose normal geometry is the cotangent bundle or a rank 2 vector bundle with the same total degree.  The ${\cal N}=2$ case was systematically studied beginning with the work \cite{Gaiotto:2009we} and the ${\cal N}=1$ versions in follow up work \cite{Bah:2012dg}.  It is natural to look for the far bigger class of 4d theories one obtains by compactifying the ${\cal N}=(1,0)$ theories.  Compactifying these theories on $T^2$ gives ${\cal N}=2$ theories \cite{Ohmori:2015pua,Ohmori:2015pia,DelZotto:2015rca}. It was shown in \cite{DelZotto:2015rca} that there is more than one non-trivial CFT one may obtain from a given theory in 6d.  Moreover, for a given theory, the end point may or may not include the moduli of $T^2$ as a moduli of the SCFT.  For some 6d theories, it was shown that there is no way to obtain a 4d SCFT whose moduli space includes the moduli of $T^2$.  The examples which lead to ${\cal N}=2$ SCFTs in 4d which have $\tau$ as a moduli space seemed to arise as F-theory geometries where in the base there is an ADE singularity.  The main aim of this paper is to extend this observation to a criterion of which (1,0) theories compactified on $T^2$
lead to ${\cal N}=2$ theories where $\tau$ appears as a moduli and thus enjoy $SL(2,\mathbb{Z})$ duality symmetry.  We then
use this to argue that for this class of theories if we consider a more general Riemann surface $\Sigma$ instead of $T^2$ we would get an ${\cal N}=1$ theory whose moduli space will include at least that of the Riemann surface moduli.  Moreover in some cases 
(such as the theory of N M5 branes probing an ADE singularity) we propose a natural candidate for the ${\cal N}=1$ moduli space.  
More recently constructions with ${\cal N}=3$ SCFT in 4d were proposed which are naturally constructed using F-theory \cite{Garcia-Etxebarria:2015wns,Aharony:2016kai}.  We show how our ${\cal N}=1$ theories can couple to these theories.

\section{Necessary Condition for $\tau$  submoduli of 4d SCFT}
It was found in \cite{DelZotto:2015rca} that if one compactifies ${\cal N}=(1,0)$ theories on $T^2$ the moduli space of $T^2$ can show up as the moduli space of the resulting 4d ${\cal N}=2$ SCFT only in special cases.  Consider for example the small $E_8$ instanton (1,0) SCFT (which in F-theory
is realized as elliptic threefold with base having an $O(-1)$ bundle over $\mathbb{P}^1$).  Compactifying
on $T^2$ we get the Minahan-Nemeschansky theory with $E_8$ global symmetry.  This theory
has no moduli and so the $\tau$ of the torus does not show up as the moduli of the 4d theory.   However cases were found in \cite{DelZotto:2015rca} which admitted a $\tau$ dependent moduli space in 4d.
It was noted there that such a dependence in F-theory construction of the 6d theories seemed to require having in the base of F-theory an ADE singularity.  For example, consider the $(1,0)$ theories which admit a single tensor branch whose F-theory base is $\mathbb{C}^2/\mathbb{Z}_{k}$, where $\mathbb{Z}_k$ acts as 
$$(z_1,z_2)\rightarrow \alpha (z_1,z_2)$$ 
with $\alpha^k=1$ (where $k\leq 12$ and $k\not= 9,10,11)$.  It was found that the 4d theory can have a $\tau$-dependent moduli space when $k$ is even, which corresponds to having
an $A_1$ singularity in the base of F-theory.\footnote{More precisely,
the F-theory base is an orbifold of an $A_1$ singularity.}
Moreover, it was shown for the case of $k=3$ that no matter how we take the limit of going down to 4 dimensions on $T^2$, the modulus of $T^2$ will not survive
as a moduli of the 4d theory.  In this case, the base is not an orbifold
of an ADE singularity.

We now argue why having an (orbifold of an) ADE type singularity in the base of F-theory is a necessary requirement for the $\tau$-dependence to show up as the moduli of the 4d ${\cal N}=2$ theory.

\subsection{Connections with Type IIA}
Consider F-theory on an elliptic 3-fold, compactified on a $T^2$ with complex structure $\tau$.  This is dual to type IIA compactified on the same elliptic 3-fold to four dimensions.  In this map the complex structure $\tau$
of $T^2$ gets mapped to the K\"ahler class of the elliptic fiber of the IIA theory.  We now ask what is the condition for the resulting 4d theory
to lead to a conformal theory for which $\tau$ is a marginal operator?  For this to happen, in the type IIA setup we need to have a situation where
the K\"ahler class of the elliptic fiber can vary arbitrarily and still lead to a conformal theory.  For this to be the case, the degrees of freedom
leading to a conformal theory, which should come from the singularity of the geometry, should include singular loci which can feel the K\"ahler class of $T^2$.  For this to be the case, the elliptic fiber must be part of the singularity locus.  This in turn means that for the elliptic threefold the singularity must include a codimension 1 space.
But singularities of CY 3-folds which have a 1 dimensional locus, must be in turn local singularity of a CY 2-folds, which in turn means that it is an ADE singularity.
Indeed this was the only class of 6d (1,0) theories which did give rise, 
upon suitable compactification on $T^2$ to conformal theories 
in 4d \cite{DelZotto:2015rca}.
(As observed in \cite{DelZotto:2015rca}, orbifolds of ADE singularities
also have this property.)
We now see that this is indeed a requirement for getting a conformal theory in 4d with ${\cal N}=2$ where $\tau$ survives as a modulus.  It is natural to ask if this is sufficient.  That this should indeed be sufficient is strongly suggested from noting that this geometry will have an N=4 subsector coming from the ADE singularity which will depend on $\tau$.  Even though this is not strictly a proof it is a plausible argument and indeed is consistent
with the findings in \cite{DelZotto:2015rca}.

\subsection{Examples} \label{subsec:examples}
In this section we give some examples which were discussed in \cite{DelZotto:2015rca}, of 6d (1,0) theories which lead upon toroidal compactification to N=2 4d SCFTs whose
moduli space includes $\tau$.

Consider an $A_{N-1}$ singularity in F-theory base.  If the elliptic fibration is trivial, compactification of this theory
on $T^2$ leads to ${\cal N}=4$ SYM in 4 dimensions, for which $\tau$ plays the role of a marginal coupling constant.
A simple ${\cal N}=(1,0)$ version of this 6d theory is to dress it up the fiber so that it has non-trivial 7-branes wrapping
the cycles of the base.  The simplest one corresponds to $I_k$ type fibers which means wrapping $k$ D7 branes over each vanishing
cycle of the base.  More generally we can consider 7-brane fiber types which lead to D and E gauge symmetries.  Let us denote the corresponding ADE gauge factors by $G$.  At the conformal point this theory has $G\times G$ global symmetry and is dual to N M5 branes probing the $G$
singularity in M-theory.  Once we compactify this theory on $T^2$ to 4d we have two options:
Not to turn on fugacities for global symmetry $G\times G$, or turn some fugacities on.
If we turn off all the fugacities for the flavor group and take $T^2$ area to zero size, as has been
argued in \cite{Ohmori:2015pua,Ohmori:2015pia} one obtains the 4d ${\cal N}=2$ theory
which is equivalent to a class S theory of $G$ type on a sphere with $N$ simple punctures and two
full punctures, leading to $G\times G$ global symmetry in 4d.  On the other hand it was also argued in
\cite{DelZotto:2015rca} that if we turn on fugacities for the diagonal flavor symmetry $G_D\subset G\times G$, we end up instead with a different 4d theory:  the affine G quiver theory with gauge group given
by 
$$\prod SU(Nd_i)$$
 where the $d_i$ are the Dynkin indices, and bifundamental matter dictated by
the links of the affine quiver.  The easiest way to see this is to note that compactifying M-theory on $T^2$
is dual to type IIB on a circle, where $M5$ brane wrapping $T^2$ becomes dual to $D3$ brane.  We thus end up with a geometry involving $N$ D3 branes probing the G=ADE singularity, leading to ${\cal N}=2$ affine quiver theory.  The moduli space of this theory is well known to be that of flat $G$ connections on $T^2$,
which in particular depends on $\tau$ (for the $A$ case see \cite{Witten:1997sc} and for the $D,E$ see \cite{Katz:1997eq}). This moduli space is identified with the choice of modulus $\tau$ as well
as choice of the flat holonomy $G_D \subset G\times G$, giving a 6d geometric explanation of the
origin of the moduli space of this 4d theory.  Turning on fugacities for the rest of the 6d flavor symmetries translates to giving masses to bifundamental fields, taking us away from the conformal fixed point.
  Note that this is an example where the geometric moduli of $T^2$ is only a subset
of moduli of the theory and the choice of flat connections on it is an added ingredient.

A concrete example of the above class can be realized as follows: we start with F-theory on elliptic 3-fold 
\begin{equation}T^2\times \mathbb{C}\times \mathbb{C}\label{eq:3fold}\end{equation}
We take the $T^2$ to have $\mathbb{Z}_3$ symmetry
and mod out this geometry  by $\mathbb{Z}_{3}\times \mathbb{Z}_{3N}$ consisting of elements:
$$(\omega^a,\omega^a \zeta^b, \omega^a \zeta^{-b})$$
where the above denotes the action on the three directions $T^2\times \mathbb{C}^2\times \mathbb{C}^2$ respectively,
and $\omega$ is a primitive third root of unity and $\zeta$ is a 
primitive $3N$-th root of unity.  This theory in the base of F-theory
has an $A_{3N-1}$ singularity which is further modded out by $\mathbb{Z}_3$ acting on the full space.  This theory is dual to $N$ M5 branes
probing an $E_6$ singularity in M-theory \cite{DelZotto:2015rca}.  Upon compactification on $T^2$ the theory leads to an ${\cal N}=2$ theory whose
moduli space is the moduli space of flat $E_6$ bundles on $T^2$ which again includes the geometric moduli $\tau$ of the torus.

As another example, also described in section 3 of \cite{DelZotto:2015rca},
we consider the E-string theory compactified to 4d
on $T^2$.  As is well-known \cite{Ganor:1996pc}, by choosing appropriate
Wilson lines we deform the $E_8$ boundary brane to two $SO(8)$ branes,
producing the $\mathcal{N}=2$, 4d $SU(2)$  gauge theory
with $N_f=4$ flavors.  The dimensionless coupling constant $\tau$ of
the theory is identified with the complex structure on $T^2$.  

The F-theory realization of this construction again starts with the elliptic
3-fold \eqref{eq:3fold} with arbitrary $T^2$, and mod out this geometry
by $\mathbb{Z}_2\times \mathbb{Z}_2$ consisting of elements
$$ ((-1)^a,(-1)^{b},(-1)^{a+b}).$$
There are three intermediate quotients where we mod out by a single
$\mathbb{Z}_2$:  the case with $b=0$ gives one of the $SO(8)$ branes
and the case with $a=b$ gives the other $SO(8)$ brane.  However, the 
$\mathbb{Z}_2$ with $a=0$ gives a base with an $A_1$ singularity,
and the final model is an additional $\mathbb{Z}_2$ quotient of this,
as expected.

\section{The General Construction}
The general idea of construction is to consider the base geometry of the CY to be the orbifold of an ADE singularity.  Since we can always
assume that the base of F-theory is an orbifold geometry 
\cite{Heckman:2013pva}  we can in principle classify all the bases which are 
an orbifold which contain an ADE subgroup.
This includes a somewhat larger class of groups than those in \cite{Heckman:2013pva} as that class was the minimal choice of the group, consistent
with the base singularity.\footnote{It
 should be possible to combine the known classification of finite
subgroups of $U(2)$ \cite{Shephard-Todd,DuVal,Coxeter,Falbel-Paupert}
with an F-theory analysis to give a complete list.
In practice, we have carried this out for what we believe to be a
complete list of cases that occur in F-theory, although we have not
done the group theory exercise needed to completely eliminate other
cases.}  We will now explain how this construction works.

We start with $\mathbb{C}^2$ and mod out by a discrete group $\Gamma$ where $\Gamma \subset U(2)$.  Moreover there is a map from
$\Gamma \rightarrow U(1)$ given by
$$\gamma\rightarrow \det(\gamma)$$
Let $H$ be the kernel of this map, which is a subgroup of $\Gamma$.  By definition $H$ will be identified with an ADE subgroup of $SU(2)$.
As already discussed we need $H$ to be non-trivial if the geometric moduli of the $T^2$ are to show up as moduli of the ${\cal N}=1$ theory.
We will perform our quotient in two steps:  first, we consider
$\mathbb{C}^2/H$ which has an ADE singularity and resolve the singularity
to obtain a space $\widetilde{\mathbb{C}^2/H}$.  The remaining group 
$G:=\Gamma/H$ lifts to an action on this space, and we then consider
the quotient by that lifted action.

In general, to build an SCFT from F-theory, we
 contract  a connected
collection of curves in the F-theory base to a point.  
It was established in \cite{Heckman:2013pva} that
after contraction, the base $B$ is an orbifold of the form $\mathbb{C}^2/\Gamma$,
where $\Gamma\subset U(2)$ is a subgroup for which the stabilizer of any
point in $\mathbb{C}^2$ other than the origin is trivial.  As pointed
out in \cite{DelZotto:2014fia,Bertolini:2015bwa}, the associated elliptic
fibration is always an orbifold of a hypersurface:  the anticanonical
bundle of the base descends from a line bundle on $\mathbb{C}^2$
on which $\gamma\in\Gamma$ acts by $\det( \gamma)$.  The action on the ``$x$'' and
``$y$'' variables used in a Weierstrass model is therefore via
$\det( \gamma)^{2}$ and $\det( \gamma)^3$, and there is a minimal Weierstrass equation
over $\mathbb{C}^2$ of the form
\[ y^2 = x^3 + f(s,t)x + g(s,t)\]
which transforms as $\det( \gamma)^{6}$.  The coefficient $g$ also transforms
as  $\det( \gamma)^{6}$, while the coefficient $f$ transforms as
 $\det( \gamma)^{4}$.  The holomorphic three-form on the hypersurface is the Poincar\'e
residue of
\[ \frac{ds\wedge dt \wedge dx \wedge dy}{-y^2+x^3+f(s,t)x+g(s,t)}\ ,\]
which is invariant under the group.

We can extend this construction by allowing for a similar quotient by
any $\Gamma\subset U(2)$, without imposing the condition about
stabilizers.  (In fact, as we shall see below, when we perform
the quotient  by $\Gamma/H$ in the second step of our basic construction,
we encounter points with nontrivial stabilizers even if we had
initially avoided them.)  For a more general quotient, we still must
act on $x$ and $y$ by $\det(\gamma)^2$ and $\det(\gamma)^3$ in order
to preserve the holomorphic three-form.

Suppose $\gamma\in\Gamma$ fixes some points other than the origin.
In this case, the images of the fixed points will
form a non-compact curve on the quotient
which support an F-theory brane representing
a flavor symmetry of the theory, with the flavor group determined
by the Kodaira type (or Kodaira--Tate type \cite{Bershadsky:1996nh}) 
of the F-theory 
brane.
By changing coordinates (and changing the generator of the cyclic
group generated by $\gamma$, if necessary), we may assume that $\gamma$
acts as $(s,t)\mapsto (e^{2\pi i/k} s, t)$ so that $s=0$ is the fixed locus.
For any fixed value of $t$, the Weierstrass equation
takes the form
\[ y^2 = x^3 + f(s)x + g(s)\]
where $\gamma$ acts on $f(s)$ by $e^{8\pi i/k}$, and $\gamma$ acts on $g(s)$
by $e^{12\pi i/k}$.

We claim that the order $k$ of $\gamma$ is at most $6$ in this situation.  
If $k>6$,
then $s^4$ is the minimum degree monomial on which
$\gamma$ acts by $e^{8\pi i/k}$, and $s^6$ is the minimum degree monomial
on which $\gamma$ acts by $e^{12\pi i/k}$.  This means that $s^4$ divides
$f(s)$ and $s^6$ divides $g(s)$.  But this is impossible for a minimal
Weierstrass model.

\begin{figure}[t]
\begin{center}
\begin{tabular}{l|rrrrr}
Group &
\multicolumn{1}{c}{$\mathbb{Z}_2$} &
\multicolumn{1}{c}{$\mathbb{Z}_3$} &
\multicolumn{1}{c}{$\mathbb{Z}_4$} &
\multicolumn{1}{c}{$\mathbb{Z}_5$} &
\multicolumn{1}{c}{$\mathbb{Z}_6$} \\ \hline
Singularities &
\multicolumn{1}{c}{$4A_1$} &
\multicolumn{1}{c}{$3A_2$} &
\multicolumn{1}{c}{$2A_3+A_1$} &
\multicolumn{1}{c}{$2A_4$} &
\multicolumn{1}{c}{$A_5+A_2+A_1$} \\ \hline
\rule{0pt}{1.65in}\raisebox{.75in}{Resolution} &
\includegraphics[scale=.55]{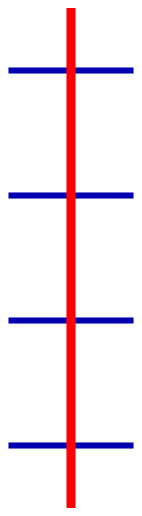} &
\includegraphics[scale=.55]{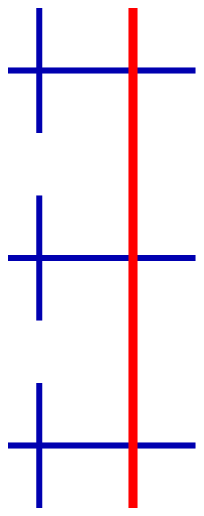}&
\includegraphics[scale=.55]{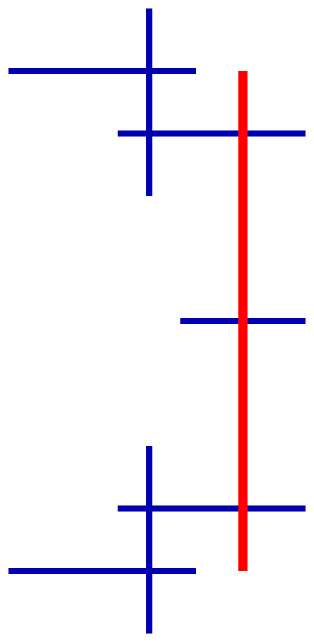}&
\includegraphics[scale=.55]{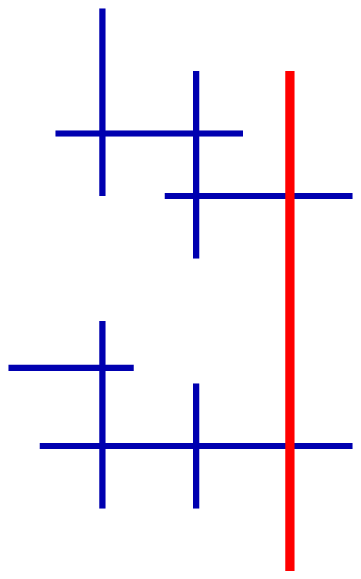}&
\includegraphics[scale=.55]{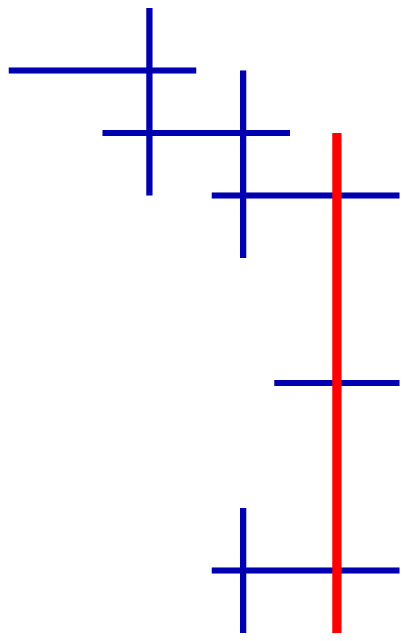}\\ \hline
Kodaira type & 
\multicolumn{1}{c}{$I_0^*$} &
\multicolumn{1}{c}{$IV^*$} &
\multicolumn{1}{c}{$III^*$} &
\multicolumn{1}{c}{$II^*$} &
\multicolumn{1}{c}{$II^*$} \\ 
\end{tabular}
\caption{Quotients along curves of fixed points.
The long vertical curve is the quotient before resolution.}\label{fig:ellipmult}
\end{center}
\end{figure}

In fact, each of the cases $k=2$, $3$, $4$, $5$, and $6$ occurs.  In
the cases $k=2$, $3$, $4$, or $6$, there is an action of the corresponding
cyclic group on $T^2$ with 3 or 4 points having nontrivial stabilizer
(as discussed earlier in this paper).
In a one-parameter family, say with $t$ constant but nonzero, the
quotient will have an $A$ type singularity at each of the fixed points.
Resolving those singularities produces a Kodaira fiber of type $I_0^*$,
$IV^*$, $III^*$, or $II^*$ (and corresponding flavor symmetry)
as illustrated in Figure~\ref{fig:ellipmult}.

The case $k=5$ is less familiar (although explained in some detail in
section 4.3 of \cite{deBoer:2001wca}).  In this case, $s^4$ divides $f(s)$
and $s$ divides $g(s)$ so the generic fiber along $s=0$  has Kodaira type $II$,
in other words, has a cusp singularity.  The group $\mathbb{Z}_5$ acts
on this with two fixed points: the ``point at infinity'', and the cuspidal
point.  The quotient gets an $A_4$ singularity at each fixed point, and when
we resolve, we find a second realization of Kodaira type $II^*$ (also
shown in Figure~\ref{fig:ellipmult}).

\subsection{Quotients of A-type}
\label{subsec:A}

We return now to the general situation of an arbitrary action of $\Gamma\subset
U(2)$ on $\mathbb{C}^2$, with $H$ the kernel of the determinant map.
We first consider the case in which the
kernel $H$ is cyclic (leading to an $A_{m-1}$ singularity on the
quotient, where $m$ is the order of $H$); other cases will be considered
in Section~\ref{subsec:D}.
 As above, we denote by $(s,t)$ the coordinates on
$\mathbb{C}^2$.  Then the resolved $A_{m-1}$ singularity 
$\widetilde{\mathbb{C}^2/H}$ can be described using $m$ coordinate
charts $W_j$, $j=0,\dots,m-1$ with coordinates $(u_j,v_j)$, which
are determined from $(s,t)$ by
\begin{align*}
u_j &= s^{m-j}/t^j \\
v_j &= t^{j+1}/s^{m-j-1} .
%u_j &= \frac{s^{m-j}}{t^j} \\
%v_j &= \frac{t^{j+1}}{s^{m-j-1}} .
\end{align*}
The change of coordinates is given by
\begin{equation}
\begin{aligned}
u_{j+1} &= 1/v_j \\
v_{j+1} &= u_jv_j^2 .
\end{aligned}
\label{eq:coordchange}
\end{equation}
Note that $u_{j+1}v_{j+1}=u_jv_j=st$.

The exceptional curves $C_j$, $j=1,\dots,m-1$ are described by
\[
C_j := \{u_j=0\} \cup \{ v_{j-1}=0\}.
\]
There are also two non-compact curves $C_0 := \{u_0=0\}$ and
$C_m := \{ v_{m-1}=0\}$.

We have a cyclic group $G=\Gamma/H$ of order $k$ acting on this space
whose determinant map is injective.  
There are two possibilities: either $G$ preserves each curve $C_j$,
or some element of $G$ maps $C_j$ to $C_{m-j}$.  We consider
the first case here, and postpone the second case to Section~\ref{subsec:DH}.

If we specify the group action 
on one of the coordinate charts then the actions on the other charts
are determined by the changes of coordinates; moreover, the action of
the determinant coincides with the action on $u_jv_j$ so it is the
same on all charts.  (It also acts on the product $st$ of the original
variables in the same way, so this is the same determinant occurring in the
$U(2)$ action.)  We choose a generator so that the determinant
acts by $e^{2\pi i/k}$.

\begin{figure}
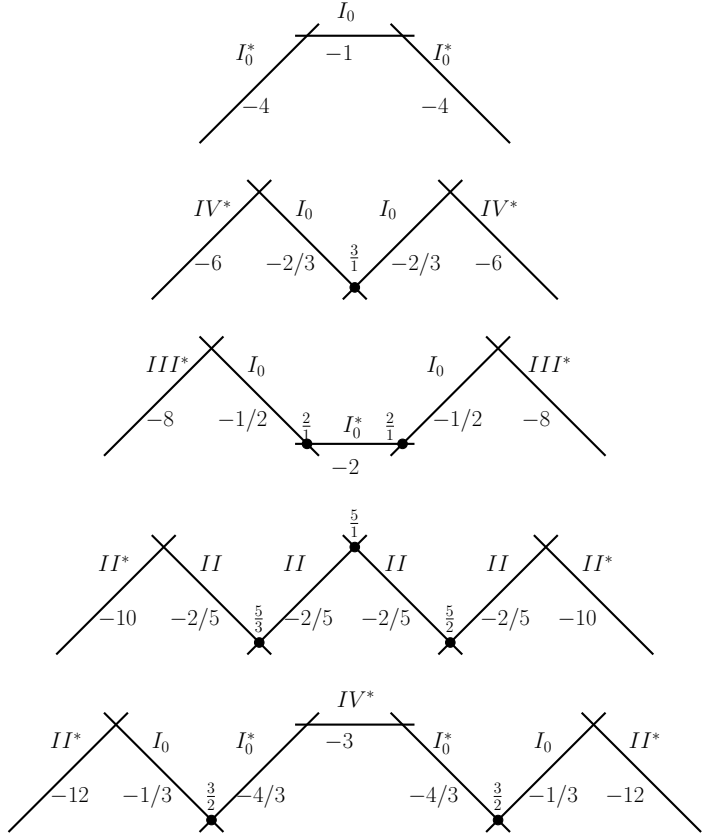


\begin{center}
\includegraphics[scale=0.5]{two-a.mps}

\bigskip

\includegraphics[scale=0.5]{three-a.mps}

\bigskip

\includegraphics[scale=0.5]{four-a.mps}

\bigskip

\includegraphics[scale=0.5]{five-a.mps}

\bigskip

\includegraphics[scale=0.5]{six-a.mps}
\end{center}
\caption{The quotients $A_{k+1}/\mathbb{Z}_k$, $k= 2, 3, 4, 5, 6$.}
\label{fig:quotients}
\end{figure}

Suppose the generator acts on $(u_0,v_0)$ by multiplication by
$(e^{2 \pi i a/k}, e^{2\pi i b/k})$.  Then $a + b \equiv 1$ mod $k$, and it
is convenient to describe the action on $(u_0,v_0)$
by the pair of rational numbers
$(\frac{a}k,\frac{1-a}k)$.  It is then easy to write down the action
on all of the charts $W_0$, $W_1$, \dots, $W_{m-1}$, using 
\eqref{eq:coordchange}:
\begin{equation}
 (\frac{a}k,\frac {1-a}k), (\frac{a-1}k,\frac{2-a}k), (\frac{a-2}k,\frac{3-a}k), \dots, (\frac{a-m+1}k,\frac{m-a}k).
\label{eq:cycle}
\end{equation}
From this we see that 
the order of the stabilizer of $C_j$ is $\gcd(j-a,k)$.  In particular, 
if $m+1\ge k$ then at least one curve $C_j$ is fixed by the $\mathbb{Z}_k$
action, and this implies that $k\le6$ (using the analysis illustrated in
Figure~\ref{fig:ellipmult}).  In Appendix~\ref{app:Atype}, 
we have analyzed
all of the F-theory bases  of A-type from \cite{Heckman:2013pva},
determining which ones are quotients of $A_{m-1}$ by $\mathbb{Z}_k$.

We can use the description in \eqref{eq:cycle} to analyze the
quotient process.  As \eqref{eq:cycle} shows, the actions on the
coordinate charts $W_j$  repeat cyclically, with the same action
on $W_j$ and $W_{j+k}$.
We will do the analysis explicitly for $m=k+2$ using $a=1$; 
any other case can be obtained by cyclically repeating this case and
truncating the ends appropriately.

We begin, therefore, with curves $C_1$, \dots, $C_{k+1}$ of
self-intersection $-2$ using the group action specified by \eqref{eq:cycle}
with $a=1$,
and let $n_j=\gcd(j-1,k)$ be the order
of the stabilizer of $C_j$.  On the quotient $A_{k+1}/\mathbb{Z}_k$, the
image of $C_j$ will have self-intersection $-2n_j^2/k$.  In addition,
there may be orbifold singularities (as determined by the group action)
at the images of the origins of the various coordinate charts $W_j$.
We display the resulting quotients (with self-intersection data)
for $k=2,3,4,5,6$ in Figure~\ref{fig:quotients}.  We have indicated the Kodaira fiber type of each curve on the
quotient (as determined by our analysis above). When $n_j=1$, the Kodaira type is type $I_0$ (indicating no fiber singularity)
for $k\ne5$, and type $II$ (cuspidal fiber) for $k=5$ as discussed above.  In the other cases,
we get Kodaira types $I_0^*$, $IV^*$, $III^*$, $II^*$, $II^*$ for
$n_j=2,3,4,5,6$.  We have also indicated the type of the orbifold singularities
which appear, by means of fractions $\frac{p_j}{q_j}$ which specify the group
action (and whose continued fraction expansion specifies the resolution).  
Note that the singularity represented by $\frac53$ when read in the opposite direction is represented by
$\frac52$; this is why both of those fractions occur in the Figure.

\begin{figure}
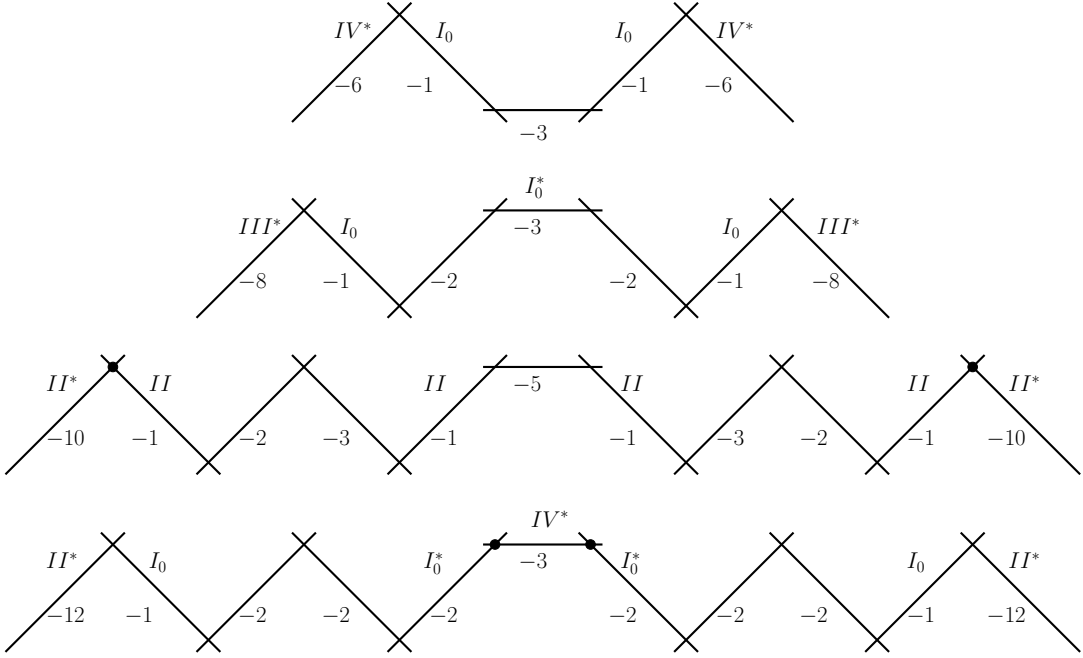


\begin{center}
\includegraphics[scale=0.5]{three-b.mps}

\bigskip

\includegraphics[scale=0.5]{four-b.mps}

\bigskip

\includegraphics[scale=0.5]{five-b.mps}

\bigskip

\includegraphics[scale=0.5]{six-b.mps}
\end{center}

\caption{The resolutions of $A_{k+1}/\mathbb{Z}_k$, $k=3,4,5,6$.}
\label{fig:resolutions}
\end{figure}

Note that the self-intersection numbers of the image curves
are sometimes fractional,
as is typically the case for surfaces with orbifold singularities
(cf.~\cite{DelZotto:2014fia}).
In Figure~\ref{fig:resolutions} we have resolved the orbifold
singularities for $k=3,4,5,6$, producing a surface with integer self-intersection
numbers.  (There are no orbifold singularities when $k=2$.)

In the cases $k=2,3,4$, this process produces  strings of the form
\begin{equation}
\begin{gathered}
4\,1\,4\cdots\\
6\,1\,3\,1\,6\cdots\\
8\,1\,232\,1\,8\cdots .\\
\end{gathered}
\end{equation}
For $k=5,6$, two further blowups are required in order to produce the
fully blown up F-theory base for the Coulomb branch:  the points to be
blown up are indicated with dots in Figure~\ref{fig:resolutions},
and after these blowups, {\em both}\/ cases yield the same string
\begin{equation}
\langle12\rangle\,1\,223\,1\,5\,1\,322\,1\,\langle12\rangle\cdots ,
\end{equation}
noting that in the case of $k=5$, a blowup is done on each end curve from 
both the left and the right, ultimately giving self-intersection $-12$
for each.
The remaining Kodaira types can be determined from the standard analysis of
non-Higgsable clusters \cite{Morrison:2012np}, but note that many of these Kodaira types
were already supplied during the quotient process.  Another interesting
feature of  cases $k=5,6$ is that each case supplies a different
subset of Kodaira types from the quotient process, but they ultimately
end up
with the same F-theory base, including the specification of Kodaira type.

The process we have described can be used no matter where the cycle starts
and stops.  In some cases, there will be curve with a non-trivial stabilizer
to the left or the right of our given curve configurations
 leading to the quotient of $\mathbb{C}^2$
by a group acting with nontrivial stabilizers; these models have
explicit global symmetries due to flavor branes to the left or the right.
This was the case for the concrete example considered in Section~\ref{subsec:examples}.

  In other cases, the curves to the left and right of the configuration
have trivial stabilizer, and the group $\Gamma$ has no 
nontrivial elements stabilizing
points of $\mathbb{C}^2$ other than the origin (although there
are fixed curves for the group action on $A_{m-1}$).  
Here,we can predict in advance some
quotients: for each value of $k$, if we truncate the cycle in such
a way that the curves to the right and the left have trivial stabilizer (i.e.,
that the numerator in the group action is relatively prime to $k$),
we will find a $\Gamma$ and a partial resolution of
$\mathbb{C}^2/\Gamma$ in the form $A_{m-1}/\mathbb{Z}_k$.  For each
$k$, the number of such possible truncations on the left is the number of congruence classes
modulo $k$ which are
relatively prime to  $k$.  This number  is one for $k=2$, two for $k=3,4,6$ and
four for $k=5$.  (There is a similar statement for truncations on the right.)
Thus there are  eleven different ways for a string to terminate
(on either side), and together with a twelfth way (``no termination other
than a string of $2$'s'') turns out to give
 a complete list of the possible terminations.
These things are explained in Appendix~\ref{app:Atype}.

Let us illustrate the construction we have made with a concrete example.
Consider $\mathbb{Z}_3\times \mathbb{Z}_m$ acting on the base of F-theory
by 
\[ (\omega^a\zeta^b, \omega^a\zeta^{-b}), \]
where $\omega$ is a primitive third root of unity and $\zeta$ is a
primitive $m$-th root of unity.
When $m=3N$,
this is the ``concrete example'' from the end of Section~\ref{subsec:examples}
(although here we have not displayed the action on the F-theory fiber,
which is determined by the action on the base).
In that case, we find that there are two curves of fixed points in
$\mathbb{C}^2$, described by $s=0$ and $t=0$ and we find an $E_6$ flavor
brane along each curve (as in the earlier description of the example).
When fully resolved, the collection of curves in between the flavor
branes corresponds to $1316\cdots6131$, which is the standard description 
of $(E_6,E_6)$ conformal matter
\cite{DelZotto:2014hpa}.

On the other hand, if $m$ is not divisible by $3$, the group is generated
by a single element, which we can take in the form $(a,b)=(-1,1)$.  After
resolving the $A_{m-1}$ singularity, the action on the chart $W_0$
is via $(\omega^{-m},\omega^{1+m})$, while if we follow the coordinate
change maps we find that the action on the chart $W_m$ is via
$(\omega^{1+m},\omega^{-m})$.  If we take the quotient and resolve
singularities, we find a curve collection of the form
$3161316\cdots613$ if $m\equiv1$ mod $3$, or a curve collection of
the form $61316\cdots61316$ if $m\equiv-1$ mod $3$.

\subsection{Quotients of D-type with $H$ cyclic} \label{subsec:DH}

We now consider the other possibility for a cyclic group acting
on an $A_{m-1}$ singularity, namely, the case in which the action reverses the order
of the curves in the $A_{m-1}$ singularity.  We assume\footnote{As indicated
in an earlier footnote, we believe that this is the general situation
but have not fully verified it.} that the number
of curves is $odd$, that is, that $m=2q$.  In addition to the action
of $\mathbb{Z}_{2q}$ on $\mathbb{C}^2$, we need another group element
which exchanges the lines $s=0$ and $t=0$ (so as to permute the
curves $C_j$).  By rescaling $s$ and
$t$ appropriately, we may assume that the action is by means of a matrix
\begin{equation}
 \lambda_r:=\begin{bmatrix} 0 & e^{2\pi i/r} \\ e^{2\pi i/r} & 0 \end{bmatrix},
\end{equation}
while the original cyclic action was by means of a matrix
\begin{equation}
 \psi_{2q}:= \begin{bmatrix} e^{2\pi i/2q} & 0 \\ 0 & e^{-2\pi i/2q}
\end{bmatrix}.
\label{eq:psi}
\end{equation}
We let $\Gamma_{q,r}$ be the group generated by $\psi_{2q}$ and 
$\lambda_r$, and investigate the quotient $\mathbb{C}^2/\Gamma_{q,r}$.

Since the action of $\lambda_r$ on $\mathbb{C}^2$ has no fixed points
other than the identity, 
the group $\Gamma_{q,r}$ is among the finite subgroups of
$U(2)$ with no nontrivial stabilized points and so the quotient
$\mathbb{C}^2/\Gamma_{q,r}$
has the potential to be an F-theory base directly
(with no flavor branes from orbifolding).  
It has a singularity of ``D-type,'' that is, one whose resolution graph
resembles $D_n$ but with different intersection numbers.
However, as we will
show in Appendix~\ref{app:Dtype}, none of these orbifolds actually
occurs as an F-theory base.

\subsection{Other quotients of D-type} \label{subsec:D}

There is another possible way to obtain a quotient of D-type:  act
on a $D_{q+2}$ singularity by a cyclic group.  The full group $\Gamma$
acting on $\mathbb{C}^2$ takes the form
\[ \mathbb{D}_{p,q}:=\langle \psi_{2q}, \tau, \phi_{2(p-q)}\rangle \]
for some integers $p>q$ with $p-q \equiv 1$ modulo 2.  
(The group takes a different form if $p-q\equiv0$ modulo 2, as explained
in Appendix~\ref{app:Dtype}.)
Here, $\psi_{2q}$
is the matrix from \eqref{eq:psi}, and the other generators are
\begin{align}
\tau &:= \begin{bmatrix} 0 & i \\ i & 0 \end{bmatrix} , \text{ and } \\
\phi_{2(p-q)} & := \begin{bmatrix} e^{2\pi i/2(p-q)} & 0 \\ 0 & 
e^{2\pi i/2(p-q)} \end{bmatrix} .
\end{align}

The subgroup with determinant $1$ is the group $\mathbb{D}_{q+1,q}$,
which leads to a $D_{q+2}$ singularity.
We begin by describing the quotient $\mathbb{C}^2/\mathbb{D}_{q+1,q}$.
Finding the functions invariant under that subgroup is a standard exercise
in invariant theory (see, for example, \cite{Slodowy}), and leads to generators
\begin{equation}
Y =\frac12( s^{2q} + t^{2q}), \quad
X = \frac12 st(s^{2q} - t^{2q}), \quad
Z = s^2t^2
\end{equation}
which satisfy the relation
\[ X^2 = Z(Y^2 - Z^q) ,\]
 the equation of a $D_{q+2}$ singularity.  The action of
$\phi_{2(p-q)}$ on the $\mathbb{D}_{q+1,q}$-invariant polynomials is
\[ (X,Y,Z) \mapsto (e^{2\pi i(q+1)/(p-q)}X,
e^{2\pi i q/(p-q)}Y, e^{4\pi i/(p-q)}Z).\]

Now we blowup $X=Z=0$ and consider the coordinate chart in which $X=WZ$.
In this chart, after substitution and 
dividing by $Z$ the equation takes the form
\[ W^2 Z = Y^2 - Z^q, \]
or,
\[ Y^2 = Z(W^2 + Z^{q-1})\]
which is a $D_{q+1}$ singularity.  The matrix $\phi_{2(p-q)}$ now acts
by
\[ (Y,W,Z) \mapsto (e^{2\pi iq/(p-q)}Y, e^{2\pi i(q-1)/(p-q)}W,
e^{4\pi i/(p-q)}Z),\]
so we recognize the remaining singular point 
as a quotient by the group $\mathbb{D}_{p-1,q-1}$.
We can thus analyze $\mathbb{C}^2/\mathbb{D}_{p,q}$ by descending induction on $q$ (keeping
$p-q$ constant).  Our blowup
has created the curve at the far end of the $D_{q+2}$ diagram,
leaving us with a $D_{q+1}$ singularity.

The exceptional divisor of the blowup
is $Z=Y=0$, and the action on that exceptional divisor is by $e^{2\pi i(q-1)/(p-q)}$
(its action on the remaining variable $W$).  We thus see a similar
phenomenon to Section~\ref{subsec:A} in which certain of the curves
in the resolved $D_{q+2}$ singularity will be stabilized by the remaining
cyclic action.

We reduce both $n$ and $q$ by further blowups until we get to $q=1$,
representing a $D_3$ singularity, which is actually an $A_3$ singularity
in different notation.  The group which acts here is 
$\mathbb{D}_{p-q+1,1}$ and since
the generator $\phi_2$ is diagonal, we can change basis so that 
\[ \tau \sim \begin{bmatrix} i & 0 \\ 0 & -i \end{bmatrix}.\]
We then recognize $\mathbb{D}_{p-q+1,q}$ as a cyclic group
of order $4(p-q)$.  The determinant of a generator is a root of unity
of order $p-q$, and the kernel of the determinant map has order $4$,
leading to the intermediate quotient $\mathbb{C}^2/\mathbb{Z}_4=A_3$.  
The next blowup of the same type produces the central curve in the $A_3$
graph (leaving $D_2=A_1+A_1$ blown down), and the determinant acts
by $e^{2\pi i(1-1)/(p-q)}$ on this central curve; in other words,
that curve is stabilized by the group.  
Thus, in order to be compatible with F-theory,
$p-q$ must be $2$, $3$, $4$, $5$, or $6$.  Since it is odd by assumption,
we have $p-q=3$ or $p-q=5$.

We claim that $p-q=5$ is impossible.  For if $p-q=5$,  then
upon taking the quotient, the central curve in the $D_{q+2}$ graph
will have Kodaira type $II^*$ (and so when the base is blown up
fully, this curve will have self-intersection $-12$).  This is not
permitted by the analysis in section 4 of \cite{Heckman:2013pva}.
Thus, $p-q=3$.

\begin{figure}
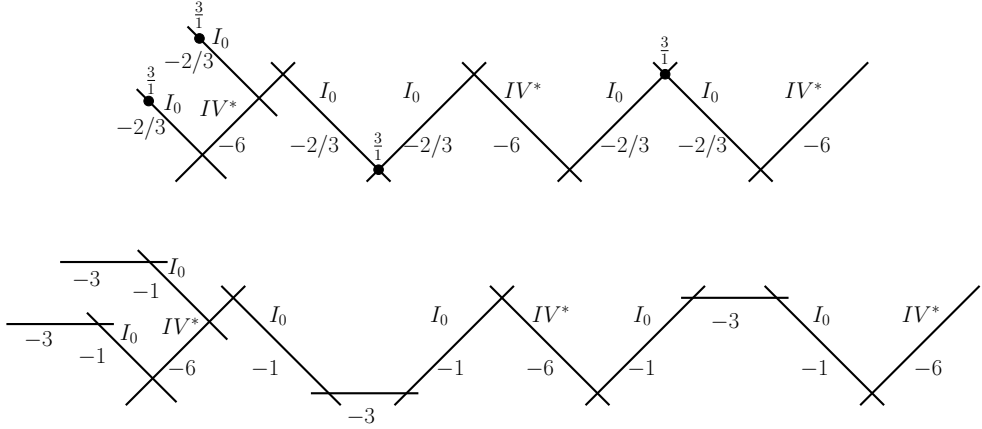


\begin{center}
\includegraphics[scale=0.5]{Dq-a.mps}

\bigskip

\bigskip

\includegraphics[scale=0.5]{Dq-b.mps}

\end{center}

\caption{The quotient $D_{q+2}/\mathbb{Z}_3$ and its resolution.}
\label{fig:Dq}
\end{figure}

In the case of $p-q=3$ we find a similar situation to the analysis
in Section~\ref{subsec:A}: the long arm of the $D_{q+2}$ graph is
acted on by $\mathbb{Z}_3$ according to the pattern \eqref{eq:cycle},
with a fixed curve at the trivalent vertex of the graph.  We display 
the corresponding quotient in Figure~\ref{fig:Dq} as well as its
resolution.  There are three cases, depending on where we truncate the
cyclic chain.  If the curve to the right of the chain is not a flavor
brane, we get one of the standard F-theory bases of D-type, as explained
in Appendix~\ref{app:Dtype}.  If the curve to the right of the chain
{\em is}\/ a flavor brane, we get a new quotient, analogous to the ones
discussed in Section~\ref{subsec:examples}.
(This happens exactly when $q$ is divisible by $3$.)
In Figure~\ref{fig:Dq}, this would correspond to the curve to the far
right being an $E_6$ flavor brane.  As can be seen from the Figure, by
first contracting all of the visible $-1$ curves and then contracting
the new $-1$ curves which were created from $-3$ curves (except at the
left) we are left with a graph of type $D_\ell$, meeting the flavor brane
at the far end of the graph.

\section{Reducing the Symmetry to ${\cal N}=1$}

In the previous section we have identified the subset of 6d (1,0) theories which when compactified on $T^2$
could lead to an ${\cal N}=2$ theory in 4d whose moduli depends on $\tau$.  Here we would like
to extend this to more general compactifications to 4d leading to ${\cal N}=1$ SCFTs in 4d.

Let us consider an ${\cal N}=(1,0)$ SCFT in 6d.  This theory will have an $SU(2)_R$ symmetry.  Twisting by $U(1)\subset SU(2)_R$
we can consider compactification of this theory on a Riemann surface $\Sigma$ which preserve half of the 6d supersymmetry.  This will yield a 4d theory with ${\cal N}=1$ supersymmetry in 4d.
The IR limit of such a theory (as we shrink the area of $\Sigma$ to zero) will lead to a fixed point.  We would like to find interesting ${\cal N}=1$
SCFTs in 4d. A class of examples of this type have been studied in \cite{Gaiotto:2015usa}.

 However, as has been the main focus of this paper, we would like to have a case where the moduli of the Riemann surface is part of the moduli of the theory.  For this to happen, since the Riemann surface moduli has a limit which factorizes to tori connected by long tubes, it is thus necessary that the toroidal compactification of the theory should also depend on the complex structure of the $T^2$. Thus we land on the subclass of theories we have discussed in the last section, whose toroidal compactification leads to 4d ${\cal N}=2$ theories with $\tau $ as part of its moduli.  

A way to construct these examples in F-theory is to take the base to be $B=\mathbb{C}\times T^*\Sigma$ where we then mod out by the group
$\Gamma$ discussed before, where we identify the $\mathbb{C}^2$ of the previous construction with $\mathbb{C}\times T^*$.  

An interesting class of examples arises by compactifying the $(1,0)$ theory corresponding to
N M5 branes probing G=ADE singularity on a Riemann surface $\Sigma$.  The condition
of preserving supersymmetry allows us to turn on arbitrary flat $G\times G $ connection 
on $\Sigma$.  On the other hand, if we are interested in getting an interesting conformal theory,
we learned that turning on the fugacities only in the diagonal $G_D\subset G\times G$ was necessary.
This will still be the case for the ${\cal N}=1$ case, if we want the theory to include the moduli
of the $\Sigma$.  In particular if we go to a limit in moduli where the Riemann surface degenerates
to tori connected to long tubes, we already know that nontrivial CFT which depends on the moduli
$\tau_i$ of the tori, arise only when we turn on  fugacity in $G_D$.  Therefore it is natural to expect
that the same is true for the ${\cal N}=1$ case. Namely the moduli space of the superconformal
theory is the moduli space of flat $G=ADE$ connections on $\Sigma$.  Note that the dimension
of this moduli space is given by
$$d=3(g-1)+\dim(ADE)(g-1)$$
and can be large as it grows not with the rank, but with the dimension of 
ADE.
It is likely that none of these theories have a Lagrangian description.  This resonates with the
results in \cite{Gaiotto:2015usa} which studies the A-type theories on a sphere with punctures
and for which the typical dual description with full punctures is believed not to be a Lagrangian theory.

As a simple example of the above class consider the theory of $N$ M5 branes probing $E_6$ singularity compactified on $\Sigma$.  In the F-theory realization this can be realized by the CY 4-fold given by orbifold of 
$$T^2\times \mathbb{C}\times T^*\Sigma/\Gamma$$
where as before $\Gamma$ has elements of the form
$$(\omega^a,\omega^a \zeta^b; \omega^a \zeta^{-b},1)$$
acting non-trivially on the $T^2\times \mathbb{C}\times T^*$ part of the geometry. 

We can also couple the above ${\cal N}=1$ theories to the recently constructed ${\cal N}=3$ theories
in \cite{Garcia-Etxebarria:2015wns}:  Consider a Riemann surface $\Sigma$ which has a $\mathbb{Z}_3$ symmetry.
Denote this action on $\Sigma$ by $\rho$.  Extend this to an action on $T^*\Sigma$ in a canonical
way which we will denote by $({\hat \rho}, \rho)$.  This in particular means near fixed points it is given by $(\omega, \omega^{-1})$.
Now consider the F-theory background on the four fold given above modded out by
an additional generator given by
$$(\omega, \omega^{-1}; {\hat \rho},{\rho})$$
For each fixed point $p_i$ of $\rho$ on the $\Sigma$ we can introduce $N_i$ D3 branes sitting
at those points.  This will realize, as has been proposed in \cite{Garcia-Etxebarria:2015wns}
a product of ${\cal N}=3$ systems for each $p_i$.  Moreover this will be coupled to the rest
of the ${\cal N}=1$ system.  So in this way we have constructed a non-trivial coupling between
these theories.  We can also consider a case when $\Sigma= T^2$ with $\mathbb{Z}_3$ symmetry,
in which case we would be getting the coupling of the ${\cal N}=2$ affine $E_6$ quiver theory to three ${\cal N}=3$ theories.
Clearly these examples can be extended to the other cases studied in \cite{Garcia-Etxebarria:2015wns}.

It is clear that we have found a rich class of ${\cal N}=1$ theories constructed from compactification
of $(1,0)$ theories in 6d.  We have only scratched the surface of this vast subject.  In particular
we expect there to be interesting types of punctures in these theories, as is known in the
context of ${\cal N}=2$ theories of class S (see in particular \cite{Gaiotto:2015usa}).

\vskip 0.5cm

\hspace*{-0.8cm} {\bf\large Acknowledgements}

\vskip 0.2cm
We would like to thank S. Razamat for valuable discussions.
CV would also like to thank KITP for hospitality during the course of this project.
This research was supported in part by the National Science Foundation
under grants PHY-1067976, PHY-1125915, and PHY-1307513.

\appendix

\section{F-theory bases of A-type}
\label{app:Atype}

In \cite{Heckman:2013pva}, the F-theory bases for all  SCFTs which can be
constructed via F-theory were classified in terms of a finite
group $\Gamma\subset U(2)$, acting without fixed points other than
the origin,  with $\mathbb{C}^2/\Gamma$ describing the F-theory base of
the SCFT.  The minimal resolution of singularities of $\mathbb{C}^2/\Gamma$
is a neighborhood of a collection of curves called an ``endpoint configuration''
in \cite{Heckman:2013pva}, since it is the endpoint of a sequence of
blowdowns from the actual F-theory base on the Coulomb branch.  In this
Appendix, we consider 
endpoint configurations of A-type, which corresponds to the group
$\Gamma$ being cyclic; the collection of curves $C_1$, \dots, $C_\ell$
forms a linear chain.  The other primary case (``D-type'') is considered
in Appendix~\ref{app:Dtype}.

The generator of the cyclic group acts on $\mathbb{C}^2$ via
\begin{equation}
(s,t) \mapsto (e^{2\pi i/p}s, e^{2\pi i q/p}t)
\end{equation}
and the continued fraction expansion of $p/q$ in the form
\begin{equation}
\frac pq = a_1 - \frac1{a_2-\frac1{a_3 - \frac{1}{\ddots}}}
\end{equation}
determines the length of the chain and the self-intersections $C_j^2=-a_j$.
We label the chain either by $p/q$ or by the string $a_1a_2a_3\cdots$
of continued fraction coefficients.

\begin{table}

%upper triangular

\begin{center}

%ut-copy1
\noindent
\begin{tabular}{c|cccccccccccc}
%headers
\diagbox[height=.8cm]{ $\alpha$}{\raisebox{.3cm}{ $\beta$}} & 
{ $7$ } & %1
{ $32222$ } & %11
$6$ & %2
{ $33$ } & %5
{ $42$ } & %7
{ $3222$ }  %10
\\ \hline
%row 1
{ $7$} &
$\frac{36N-24}{6N-5}$ & %1
$\frac{36N-144}{6N-25}$ & %11
$\frac{30N-19}{5N-4}$ & %2
$\frac{30N-37}{5N-7}$ & %5
$\frac{30N-43}{5N-8}$ & %7
$\frac{30N-91}{5N-16}$  %10
\\
%row 11
{$22223$} & 
& %1
$\frac{36N-264}{30N-221}$& %11
$\frac{30N-119}{25N-100}$& %2
$\frac{30N-137}{25N-115}$& %5
$\frac{30N-143}{25N-120}$& %7
$\frac{30N-191}{25N-160}$ %10
\\
%row 2
$6$ &
& %1
& %11
$\frac{25N-15}{5N-4}$ & %2
$\frac{25N-30}{5N-7}$ & %5
$\frac{25N-35}{5N-8}$ & %7
$\frac{25N-75}{5N-16}$  %10
\\
%row 5
{$33$}  & 
& %1 
& %11
& %2
$\frac{25N-45}{10N-19}$ & %5
$\frac{25N-50}{10N-21}$ & %7
$\frac{25N-90}{10N-37}$ %10
\\
%row 7
{$24$}  & 
& %1
& %11
& %2
& %5
$\frac{25N-55}{15N-34}$ & %7
$\frac{25N-95}{15N-58}$ %10
\\
%row 10
{$2223$} & 
& %1
& %11
& %2
& %5
& %7
$\frac{25N-135}{20N-109}$ %10
\\
\end{tabular}

\smallskip

%ut-copy2
\noindent
\begin{tabular}{c|cccccccccccc}
%headers
\diagbox[height=.8cm]{ $\alpha$}{\raisebox{.3cm}{ $\beta$}} & 
$5$ & %3
{ $322$ } & %9
$4$ & %4
$32$ & %8
$3$ & %6
$\emptyset$ %12
\\ \hline
%row 1
{ $7$} &
$\frac{24N-14}{4N-3}$ & %3
$\frac{24N-50}{4N-9}$ & %9
$\frac{18N-9}{3N-2}$ & %4
$\frac{18N-21}{3N-4}$ & %8
$\frac{12N-4}{2N-1}$ & %6
$\frac{6N+1}{N}$ %12
\\
%row 11
{$22223$} & 
$\frac{24N-94}{20N-79}$& %3
$\frac{24N-130}{20N-109}$& %9
$\frac{18N-69}{15N-58}$& %4
$\frac{18N-81}{15N-68}$& %8
$\frac{12N-44}{10N-37}$& %6
$\frac{6N-19}{5N-16}$ %12
\\
%row 2
$6$ &
$\frac{20N-11}{4N-3}$ & %3
$\frac{20N-41}{4N-9}$ & %9
$\frac{15N-7}{3N-2}$ & %4
$\frac{15N-17}{3N-4}$ & %8
$\frac{10N-3}{2N-1}$ & %6
$\frac{5N+1}{N}$ %12
\\
%row 5
{$33$}  & 
$\frac{20N-23}{8N-10}$ & %3
$\frac{20N-53}{8N-22}$& %9
$\frac{15N-16}{6N-7}$ & %4
$\frac{15N-26}{6N-11}$ & %8
$\frac{10N-9}{4N-4}$ & %6
$\frac{5N-2}{2N-1}$ %12
\\
%row 7
{$24$}  & 
$\frac{20N-27}{12N-17}$ & %3
$\frac{20N-57}{12N-35}$& %9
$\frac{15N-19}{9N-12}$& %4
$\frac{15N-29}{9N-18}$ & %8
$\frac{10N-11}{6N-7}$ & %6
$\frac{5N-3}{3N-2}$ %12
\\
%row 10
{$2223$} & 
$\frac{20N-59}{16N-48}$& %3
$\frac{20N-89}{16N-72}$& %9
$\frac{15N-43}{12N-35}$& %4
$\frac{15N-53}{12N-43}$& %8
$\frac{10N-27}{8N-22}$& %6
$\frac{5N-11}{4N-9}$ %12
\\
\end{tabular}

\smallskip

%ut-copy3
\noindent
\begin{tabular}{c|cccccccccccc}
%headers
\diagbox[height=.8cm]{ $\alpha$}{\raisebox{.3cm}{ $\beta$}} & 
$5$ & %3
{ $322$ } & %9
$4$ & %4
$32$ & %8
$3$ & %6
$\emptyset$ %12
\\ \hline
%row3
$5$ &
$\frac{16N-8}{4N-3}$ & %3
$\frac{16N-32}{4N-9}$ & %9
$\frac{12N-5}{3N-2}$ & %4
$\frac{12N-13}{3N-4}$ & %8
$\frac{8N-2}{2N-1}$ & %6
$\frac{4N+1}{N}$ %12
\\
%row 9
{$223$}  & 
& %3
$\frac{16N-56}{12N-43}$& %9
$\frac{12N-23}{9N-18}$& %4
$\frac{12N-31}{9N-24}$& %8
$\frac{8N-14}{6N-11}$& %6
$\frac{4N-5}{3N-4}$ %12
\\
%row 4
$4$ &
& %3
& %9
$\frac{9N-3}{3N-2}$ & %4
$\frac{9N-9}{3N-4}$ & %8
$\frac{6N-1}{2N-1}$ & %6
$\frac{3N+1}{N}$ %12
\\
%row 8
$23$ & 
& %3
& %9
& %4
$\frac{9N-15}{6N-11}$ & %8
$\frac{6N-5}{4N-4}$ & %6
$\frac{3N-1}{2N-1}$ %12
\\
%row6 
$3$ &
& %3
& %9
& %4
& %8
$\frac{4N}{2N-1}$ & %6
$\frac{2N+1}{N}$ %12
\\
%row 12
$\emptyset$ &
& %3
& %9
& %4
& %8
& %6
$\frac{N+1}{N}$ %12
\\
\end{tabular}

\end{center}

\caption{The continued fractions for endpoint configurations.}
\label{tab:contfrac}
\end{table}

It was found in \cite{Heckman:2013pva} that the possible endpoint
configurations have a regular behavior once the number of curves
is at least $10$.  The string $a_1a_2a_3\cdots a_N$ of length $N$
 takes the form $\alpha A_{N-a-b} \beta$
for certain strings $\alpha$ and $\beta$ (which may be empty), where
$A_{N-a-b}$ denotes a string of $N{-}a{-}b$ $2$'s, and
where $a$ is the number of entries in $\alpha$ and $b$ is the
number of entries in $\beta$. 
The list of possible
$\alpha$'s and $\beta$'s given in \cite{Heckman:2013pva} was
somewhat implicit:  obtaining the full list from the data given
there involves lowering certain  entries below their maximal
values (which were explicitly listed).  Here, we use the entire set
of possibilities, and find a beautiful correspondence with the
structure of quotients.  From the implicit description in
\cite{Heckman:2013pva}, one finds that there are precisely $12$ possibilities
for each of $\alpha$ and $\beta$.
The $\beta$'s are
simply the $\alpha$'s with their order reversed.

Ref.~\cite{Heckman:2013pva} also classified endpoint configurations
with $N<10$ entries, and these include all strings
 $\alpha A_{N-a-b} \beta$ with $N\ge a+b$.
Table~\ref{tab:contfrac} gives a formula for $p/q$ in
terms of $N$ for each possible pair $(\alpha,\beta)$, valid
for $N\ge a+b$.  Since reversing
the order of the chain does not affect the orbifold (although it
may affect the generator of the group), we only include an ``upper
triangular'' array of $78$ cases.  For ease of reading, we have
divided our array into three pieces, representing the second, first,
and fourth quadrants of a larger array.

We then extend the formulas from
Table~\ref{tab:contfrac} to a smaller value of $N$
(where the interpretation as $\alpha A_{N-a-b} \beta$ is lost).
In each entry of Table~\ref{tab:extrapolate},  we have evaluated the formula from
Table~\ref{tab:contfrac} for  $N=a+b-1$ (assuming $ab \ne0$)
and then expressed the result in terms of its continued
fraction coefficients.  Remarkably, the length of the string is $a+b-1$
in each case.\footnote{This fails to be true if the formula
is extrapolated to even smaller values of $N$.}
 Also remarkably, each of these extrapolated
strings is one of the endpoints of an F-theory configuration.
In fact,
the full list of F-theory endpoint configurations with $N<10$
(as classified in \cite{Heckman:2013pva}) is given by the strings
in Table~\ref{tab:extrapolate} together with all entries in
 Table~\ref{tab:contfrac} satisfying $a+b\le N<10$.  
These Tables thus represent
a compact summary and slight refinement of the classification 
given in Tables~1 and 2 of \cite{Heckman:2013pva}.\footnote{In
the course of compiling these extended Tables, we discovered that
the entry for $(\alpha,\beta)=(72222,22233)$ in Table~2 of \cite{Heckman:2013pva}
should read $\frac{30N+263}{5N+43}$.  Note that direct comparison
with \cite{Heckman:2013pva} is tricky because of our notation
changes.}

\begin{table}

%extrapolated string

\begin{center}

%ES-copy1
\noindent
\begin{tabular}{c|cccccccccccc}
%headers
\diagbox[height=.8cm]{ $\alpha$}{\raisebox{.3cm}{ $\beta$}} & 
{ $7$ } & %1
{ $32222$ } & %11
$6$ & %2
{ $33$ } & %5
{ $42$ } & %7
{ $3222$ }  %10
\\ \hline
%row 1
{ $7$} &
$\langle12\rangle$ & %1
$82222$ & %11
$\langle11\rangle$ & %2
$83$ & %5
$92$ & %7
$8222$  %10
\\
%row 11
{$22223$} & 
& %1
$222242222$& %11
$22227$& %2
$222243$& %5
$222252$& %7
$22224222$ %10
\\
%row 2
$6$ &
& %1
& %11
$\langle10\rangle$ & %2
$73$ & %5
$82$ & %7
$7222$  %10
\\
%row 5
{$33$}  & 
& %1 
& %11
& %2
$343$ & %5
$352$ & %7
$34222$ %10
\\
%row 7
{$24$}  & 
& %1
& %11
& %2
& %5
$262$ & %7
$25222$ %10
\\
%row 10
{$2223$} & 
& %1
& %11
& %2
& %5
& %7
$2224222$ %10
\\
\end{tabular}

\smallskip

%ES-copy2
\noindent
\begin{tabular}{c|cccccccccccc}
%headers
\diagbox[height=.8cm]{ $\alpha$}{\raisebox{.3cm}{ $\beta$}} & 
$5$ & %3
{ $322$ } & %9
$4$ & %4
$32$ & %8
$3$ & %6
$\emptyset$ %12
\\ \hline
%row 1
{ $7$} &
$\langle10\rangle$ & %3
$822$ & %9
$9$ & %4
$82$ & %8
$8$ & %6
-- %12
\\
%row 11
{$22223$} & 
$22226$& %3
$2222422$& %9
$22225$& %4
$222242$& %8
$22224$& %6
-- %12
\\
%row 2
$6$ &
$9$ & %3
$722$ & %9
$8$ & %4
$72$ & %8
$7$ & %6
-- %12
\\
%row 5
{$33$}  & 
$36$ & %3
$3422$& %9
$35$ & %4
$342$ & %8
$34$ & %6
-- %12
\\
%row 7
{$24$}  & 
$27$ & %3
$2522$& %9
$26$& %4
$252$ & %8
$25$ & %6
-- %12
\\
%row 10
{$2223$} & 
$2226$& %3
$222422$& %9
$2225$& %4
$22242$& %8
$2224$& %6
-- %12
\\
\end{tabular}

\smallskip

%ES-copy3
\noindent
\begin{tabular}{c|cccccccccccc}
%headers
\diagbox[height=1cm]{ $\alpha$}{\raisebox{.3cm}{ $\beta$}} & 
$5$ & %3
{ $322$ } & %9
$4$ & %4
$32$ & %8
$3$ & %6
$\emptyset$ %12
\\ \hline
%row3
$5$ &
$8$ & %3
$622$ & %9
$7$ & %4
$52$ & %8
$6$ & %6
-- %12
\\
%row 9
{$223$}  & 
& %3
$22422$& %9
$225$& %4
$2242$& %8
$224$& %6
-- %12
\\
%row 4
$4$ &
& %3
& %9
$6$ & %4
$52$ & %8
$5$ & %6
-- %12
\\
%row 8
$23$ & 
& %3
& %9
& %4
$242$ & %8
$24$ & %6
-- %12
\\
%row6 
$3$ &
& %3
& %9
& %4
& %8
$4$ & %6
-- %12
\\
%row 12
$\emptyset$ &
& %3
& %9
& %4
& %8
& %6
-- %12
\\
\end{tabular}

\end{center}

\caption{The extrapolated endpoints.  Note that $\langle10\rangle$,
$\langle11\rangle$ and $\langle12\rangle$ denote single-entry strings
with a 2-digit entry.}
\label{tab:extrapolate}
\end{table}

\begin{table}

%covering data

\begin{center}

%cd-copy1
\noindent
\begin{tabular}{c|cccccccccccc}
%headers
\diagbox[height=1cm]{ $\alpha$}{\raisebox{.3cm}{ $\beta$}} & 
{ $7$ } & %1
{ $32222$ } & %11
$6$ & %2
{ $33$ } & %5
{ $42$ } & %7
{ $3222$ }  %10
\\ \hline
%row 1
{ $7$} &
$(6N{-}4,\frac{6}{1})$ & %1
$(6N{-}24,\frac{6}{1})$ & %11
$(1,\frac{30N-19}{5N-3})$ & %2
$(1,\frac{30N-37}{5N-6})$ & %5
$(1,\frac{30N-43}{5N-7})$ & %7
$(1,\frac{30N-91}{5N-15})$  %10
\\
%row 11
{$22223$} & 
& %1
$(6N{-}44,\frac{6}{5})$& %11
$(1,\frac{30N-119}{25N-99})$& %2
$(1,\frac{30N-137}{25N-114})$& %5
$(1,\frac{30N-143}{25N-119})$& %7
$(1,\frac{30N-191}{25N-159})$ %10
\\
%row 2
$6$ &
& %1
& %11
$(5N{-}3,\frac{5}{1})$ & %2
$(5N{-}6,\frac{5}{1})$ & %5
$(5N{-}7,\frac{5}{1})$ & %7
$(5N{-}15,\frac{5}{1})$  %10
\\
%row 5
{$33$}  & 
& %1 
& %11
& %2
$(5N{-}9,\frac{5}{2})$ & %5
$(5N{-}10,\frac{5}{2})$ & %7
$(5N{-}18,\frac{5}{2})$ %10
\\
%row 7
{$24$}  & 
& %1
& %11
& %2
& %5
$(5N{-}11,\frac{5}{3})$ & %7
$(5N{-}19,\frac{5}{3})$ %10
\\
%row 10
{$2223$} & 
& %1
& %11
& %2
& %5
& %7
$(5N{-}27,\frac{5}{4})$ %10
\\
\end{tabular}

\smallskip

%cd-copy2
\noindent
\begin{tabular}{c|cccccccccccc}
%headers
\diagbox[height=1cm]{ $\alpha$}{\raisebox{.3cm}{ $\beta$}} & 
$5$ & %3
{ $322$ } & %9
$4$ & %4
$32$ & %8
$3$ & %6
$\emptyset$ %12
\\ \hline
%row 1
{ $7$} &
$(2,\frac{12N-7}{2N-1})$ & %3
$(2,\frac{12N-25}{2N-4})$ & %9
$(1,\frac{18N-9}{3N-1})^\dagger$ & %4
$(3,\frac{6N-7}{N-1})$ & %8
$(4,\frac{3N-1}{\frac N2})^*$ & %6
$(5,\frac{6(\frac{N+1}5)-1}{\frac{N+1}5})^*$ %12
\\
%row 11
{$22223$} & 
$(2,\frac{12N-47}{10N-39})$& %3
$(2,\frac{12N-65}{10N-54})$& %9
$(3,\frac{6N-23}{5N-19})$& %4
$(1,\frac{18N-81}{15N-67})^\ddagger$& %8
$(4,\frac{3N-11}{5(\frac N2)-9})^*$& %6
$(5,\frac{6(\frac{N+1}5)-5}{5(\frac{N+1}5)-4})^*$ %12
\\
%row 2
$6$ &
$(1,\frac{20N-11}{4N-2})$ & %3
$(1,\frac{20N-41}{4N-8})$ & %9
$(2,\frac{15(\frac{N+1}2)-11}{3(\frac{N+1}2)-2})^*$ & %4
$(2,\frac{15(\frac{N+1}2)-16}{3(\frac{N+1}2)-3})^*$ & %8
$(3,\frac{10(\frac{N}3)-1}{2(\frac N3)})^*$ & %6
$(4,\frac{5(\frac{N+1}4)-1}{\frac{N+1}4})^*$ %12
\\
%row 5
{$33$}  & 
$(1,\frac{20N-23}{8N-9})$ & %3
$(1,\frac{20N-53}{8N-21})$& %9
$(2,\frac{15(\frac N2)-8}{3N-3})^*$ & %4
$(2,\frac{15(\frac N2)-13}{3N-5})^*$ & %8
$(3,\frac{10(\frac N3)-3}{4(\frac N3)-1})^*$ & %6
$(4,\frac{5(\frac{N+2}4)-3}{\frac{N+2}2-1})^*$ %12
\\
%row 7
{$24$}  & 
$(1,\frac{20N-27}{12N-16})$ & %3
$(1,\frac{20N-57}{12N-34})$& %9
$(2,\frac{15(\frac{N+1}2)-17}{9(\frac{N+1}2)-10})^*$& %4
$(2,\frac{15(\frac{N+1}2-22}{9(\frac{N+1}2)-13})^*$ & %8
$(3,\frac{10(\frac{N+1}3)-7}{6(\frac{N+1}3)N-4})^*$ & %6
$(4,\frac{5(\frac{N+1}4)-2}{3(\frac{N+1}4)-1})^*$ %12
\\
%row 10
{$2223$} & 
$(1,\frac{20N-59}{16N-47})$& %3
$(1,\frac{20N-89}{16N-71})$& %9
$(2,\frac{15(\frac{N+1}2)-29}{12(\frac{N+1}2)N-23})^*$& %4
$(2,\frac{15(\frac{N+1}2)N-34}{12(\frac{N+1}2)-27})^*$& %8
$(3,\frac{10(\frac N3)-9}{8(\frac N3)-7})^*$& %6
$(4,\frac{5(\frac{N+1}4)-4}{4(\frac{N+1}4)-3})^*$ %12
\\
\end{tabular}

\smallskip

%cd-copy3
\noindent
\begin{tabular}{c|cccccccccccc}
%headers
\diagbox[height=1cm]{ $\alpha$}{\raisebox{.3cm}{ $\beta$}} & 
$5$ & %3
{ $322$ } & %9
$4$ & %4
$32$ & %8
$3$ & %6
$\emptyset$ %12
\\ \hline
%row3
$5$ &
$(4N{-}2,\frac{4}{1})$ & %3
$(4N{-}8,\frac{4}{1})$ & %9
$(1,\frac{12N-5}{3N-1})$ & %4
$(1,\frac{12N-13}{3N-3})$ & %8
$(2,\frac{4N-1}{N})$ & %6
$(3,\frac{4(\frac{N+1}3)-1}{\frac{N+1}3})^*$ %12
\\
%row 9
{$223$}  & 
& %3
$(4N{-}14,\frac{4}{3})$& %9
$(1,\frac{12N-23}{9N-17})$& %4
$(1,\frac{12N-31}{9N-23})$& %8
$(2,\frac{4N-7}{3N-5})$& %6
$(3,\frac{4(\frac{N+1}3)N-3}{3(\frac{N+1}3)-2})^*$ %12
\\
%row 4
$4$ &
& %3
& %9
$(3N{-}1,\frac{3}{1})$ & %4
$(3N{-}3,\frac{3}{1})$ & %8
$(1,\frac{6N-1}{2N})$ & %6
$(2,\frac{3(\frac{N+1}2)11}{\frac{N+1}2})^*$ %12
\\
%row 8
$23$ & 
& %3
& %9
& %4
$(3N{-}5,\frac{3}{2})$ & %8
$(1,\frac{6N-5}{4N-3})$ & %6
$(2,\frac{3(\frac{N+1}2)-2}{2(\frac{N+1}2)-1})^*$ %12
\\
%row6 
$3$ &
& %3
& %9
& %4
& %8
$(2N,\frac{2}{1})$ & %6
$(1,\frac{2N+1}{N+1})$ %12
\\
%row 12
$\emptyset$ &
& %3
& %9
& %4
& %8
& %6
$(N{+}1,\frac{1}{1})$ %12
\\
\end{tabular}

\end{center}

\caption{The data $(m,k/\ell)$ determining the structure of the kernel
of $\det$
and the intermediate quotient.  Footnotes are described in the text.}
\label{tab:quotdata}
\end{table}

In Table~\ref{tab:quotdata}, we analyze the behavior of the determinant
map.  Given $p/q$ describing the original group action, its determinant
acts by multiplication by $e^{2\pi i(q+1)/p}$.  In order to find
the kernel of this,
for each of the entries $p/q$ in Table~\ref{tab:contfrac} we write
\[ \frac p{q+1} = m \, \frac k\ell,\]
where $k$ and $\ell$ are relatively prime, and list the pair $(m,k/\ell)$
in Table~\ref{tab:quotdata}.  Then the kernel $H$ of the determinant
map $\det:\Gamma\to U(1)$ has order $m$, and the generator of
the residual group action by $\Gamma/H$ has determinant $e^{2\pi i\ell/k}$.
Note that taking the quotient by $H$ gives an intermediate
$A_{m-1}$ singularity, and the remaining quotient is by
$\mathbb{Z}_k$.

We have arranged the entries in the Tables to make the role of the
quotient construction manifest:  the rows and columns are sorted according
to which value of $k$ they correspond.  Let us indicate how the
description in Section~\ref{subsec:A} leads to the description here.
For example, in the case $k=5$ there are four possible ways of truncating
the cycle in Figure~\ref{fig:quotients} from the left (which we describe
in terms of the  resolved version illustrated
in Figure~\ref{fig:resolutions}):
\begin{equation}
\begin{aligned}
&\langle10\rangle\,123151321\,\langle10\rangle\,123151321\,\langle10\rangle
\,123151321\,\cdots\\
&321\,\langle10\rangle\,123151321\,\langle10\rangle\,123151321\,\langle10\rangle
\,123151321\,\cdots\\
&51321\,\langle10\rangle\,123151321\,\langle10\rangle\,123151321\,\langle10\rangle
\,123151321\,\cdots\\
&23151321\,\langle10\rangle\,123151321\,\langle10\rangle\,123151321\,\langle10\rangle
\,123151321\,\cdots\\
\end{aligned}
\end{equation}
These blow down to endpoint configurations
\begin{equation}
\begin{aligned}
&622\cdots\\
&2422\cdots\\
&3322\cdots\\
&222322\cdots\\
\end{aligned}
\end{equation}
which determine choices for $\alpha$ corresponding to $k=5$
(namely, $\alpha=6$, $24$, $33$, or $2223$).

Table~\ref{tab:quotdata} has a few footnotes, which we now explain.
 For some entries in the table, a congruence condition must be satisfied
in order to obtain the given value of $m$, and other values of $m$
can be inferred when the congruence condition fails.
For example,
$$(4,\frac{5(\frac{N+1}4)-1}{\frac{N+1}4})^*$$
means that $m=4$ when $N\equiv 3$ mod 4.  But when $N\equiv1$ mod 4 we have
$m=2$ and
$$(2,\frac{5(\frac{N+1}2)-2}{\frac{N+1}2})^*$$
and when $N\equiv 0$ mod 2 we have $m=1$ and
$$(1,\frac{5(N+1)-4}{N+1})^*$$

Also, for all but two entries of Table~\ref{tab:quotdata}, a fraction of the form
$$\frac{aN+k}{bN+\ell}$$
is seen to have relatively prime numerator and denominator
by determining the greatest common divisor
$c$ of $a$ and $b$, and then computing
$$\frac bc (aN+k) - \frac ac (bN+\ell)= \pm1.$$
In two cases (indicated by daggers) this does not work, and a different
method is needed to show that the numbers are relatively prime.
Here are the relations which are needed:
\begin{gather*}
{}^\dagger (6N{-}1)(3N-1) - N (18N-9) = 1 \\
{}^\ddagger (12N{-}52)(15N-67) - (10N{-}43)(18N-81) = 1
\end{gather*}

Table~\ref{tab:quotdata} exhibits the behavior 
predicted in Section~\ref{subsec:A}.  
Each $\mathbb{Z}_k$ has a set of associated strings $\alpha$ and $\beta$
which serve as the far left and far right ends of the configuration
arising from a quotient constructions.
As can be be seen from the Table, the strings $\alpha$ are sorted according
to $k$ as follows.

\begin{center}
\begin{tabular}{c|c|c|c|c|c|c}
$k$ & $6$ & $5$ & $4$ & $3$ & $2$ & $1$ \\ \hline
$\alpha$ & $7$, $22223$ & $6$, $33$, $24$, $2223$ & 
$5$, $223$ & $4$, $23$ & $3$ & $\emptyset$ \\
\end{tabular}
\end{center}

\noindent
The strings $\beta$ are simply the same strings read in reverse order.
Note that we have assigned $k=1$ (no quotient) to $\emptyset$ (no
string other than $2$'s at the far left).
Whenever $\alpha$ and $\beta$ are chosen from the same group, the
resulting orbifold is a $\mathbb{Z}_k$ quotient of an $A_{m-1}$ singularity.
These entries occur along the block-diagonal of Table~\ref{tab:quotdata}.

Apart from those block-diagonal entries, we observe that the value of
$m$ always lies between $1$ and $5$.  We do not at present
have an explanation for
that observation, 
or for the evident regularity in the values of $m$ which can
be seen in the data.

\section{F-theory bases of D-type}
\label{app:Dtype}

\begin{figure}

\begin{center}
\includegraphics[scale=0.5]{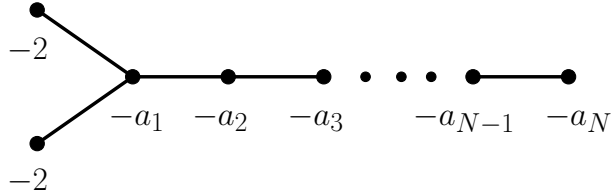}
\end{center}

\caption{Resolution graph of D-type singularity.}
\label{fig:Dtype}

\end{figure}

We now turn to the F-theory bases of D-type.  In this case, 
according to \cite{Heckman:2013pva}, the minimal
resolution of the orbifold singularity takes the form $D_{N}\gamma$,
where $\gamma\in\{ 32, 24\}$.
(Our convention this time is that the resolution graph contains
$N+2\ge4$ nodes.)
The groups which produce D-type orbifolds have been classified
\cite{Brieskorn,Riemenschneider,Iyama-Wemyss} and are closely related
to the resolution.  In fact, the resolution is expressed in terms
of a continued fraction expansion 
\begin{equation}
\frac pq = a_1 - \frac1{a_2-\frac1{a_3 - \frac{1}{\ddots}}}
\end{equation}
with the continued fraction coefficients becoming (negatives of) self-intersection numbers, as indicated in Figure~\ref{fig:Dtype}.  The corresponding group is
\begin{equation}
\mathbb{D}_{p,q} = \begin{cases} \langle \psi_{2q}, \tau, \phi_{2(p-q)}\rangle & \text{if } p-q\equiv1\mod2 \\
\langle \psi_{2q}, \tau\phi_{4(p-q)} \rangle & \text{if }p-q\equiv0\mod2
\end{cases}
\end{equation}
and has order $4q(p-q)$.
The groups appearing in Section~\ref{subsec:DH} are of type $\mathbb{D}_{p,q}$
with $p-q$ even, while the groups appearing in Section~\ref{subsec:D}
are of type $\mathbb{D}_{p,q}$ with $p-q$ odd.

We now compute the fraction $p/q$ for the cases $D_{N}\gamma$
of relevance to F-theory.  In case $\gamma=32$, the fraction is
$(3N-1)/(3N-4)$ while in the case $\gamma=24$, the fraction is
$(3N+1)/(3N-2)$.  (Both formulas are valid for $N\ge2$.)
In both cases $p-q=3$ is odd, so the even groups 
never make an appearance.  The fact that $p-q=3$ and that there are
precisely two cases of this (depending on where the quotient graph
in the top of Figure~\ref{fig:Dq}
is truncated) was expected from the analysis in Section~\ref{subsec:D}.

\end{document}